\documentclass[twocolumn,secnumarabic,amsmath,amssymb,nobibnotes,aps,pra,showkeys,superscriptaddress,floatfix,tightenlines]{revtex4-2}

\usepackage{amsfonts}
\usepackage[caption=false, font=footnotesize]{subfig}
\usepackage{graphicx}
\usepackage{wrapfig}
\usepackage[mathscr]{eucal}
\usepackage{soul}
\usepackage{tabularx}
\usepackage{makecell}
\usepackage{enumitem}
\usepackage{nicefrac}

\usepackage[rawfloats=true]{floatrow} 
\restylefloat{figure}

\usepackage[colorlinks,citecolor=blue,urlcolor=blue,bookmarks=false,hypertexnames=true]{hyperref}

\begin{document}
	
	\title{Simulating quantum field theories on gate-based quantum computers}
	
	\author{Gayathree M.~Vinod and Anil Shaji}
	\email[Corresponding author: ]{gaya3mv17@iisertvm.ac.in}
	\affiliation{School of Physics, IISER Thiruvananthapuram, Maruthamala PO, Vithura, Kerala, India}
	
	
	\begin{abstract}
		We implement a simulation of a quantum field theory in 1+1 space-time dimensions on a gate-based quantum computer using the light front formulation of the theory. The non-perturbative simulation of the Yukawa model field theory is verified on IBM's simulator and is also demonstrated on a small-scale IBM circuit-based quantum processor, on the cloud, using IBM Qiskit. The light front formulation allows for controlling the resource requirement and complexity of the computation with commensurate trade-offs in accuracy and detail by modulating a single parameter, namely the harmonic resolution. Qubit operators for the bosonic excitations were also created and were used along with the fermionic ones already available, to simulate the theory involving all of these particles. With the restriction on the number of logical qubits available on the existent gate-based Noisy Intermediate-Scale Quantum (NISQ) devices, the trotterization approximation is also used. We show that experimentally relevant quantities like cross sections for various processes, survival probabilities of various states, etc. can be computed. We also explore the inaccuracies introduced by the bounds on achievable harmonic resolution and Trotter steps placed by the limited number of qubits and circuit depth supported by present-day NISQ devices. 
	\end{abstract}
	
	
	\maketitle
	
    \section{Introduction} \label{intro}
    
    Simulation of complex quantum systems that are beyond the reach of classical computers is one of the primary roles envisaged for quantum computers, at least in the near term \cite{Feynman}. Developing methods and techniques that allow for simulation \cite{Nielsen_Chuang} of quantum many-body systems like large molecules, solid state lattices etc. on presently day Noisy Intermediate-Scale Quantum (NISQ) devices is a very active area of research \cite{NISQ_Preskill, QC_chem, QC_condensed_matter, QC_condensed_matter_2, QC_condensed_matter_3, QC_condensed_matter_4, QC_protein_folding, Preskill}. The Hilbert space of states of such a many-body system and that of a sufficiently large qubit register are virtually identical and mapping of states and operators of the real system to those that can be implemented on the quantum computer that houses the qubit register is, in a nutshell, the problem that is being addressed by the simulation methods.

    Quantum fields - unlike the quantum particles involved in many body physics - are extended objects with a Hilbert space structure that has a few crucial differences \cite{Peskin, Maggiore}. In many-body systems, typically, conservation of the number of particles is a reasonable assumption at the energy scales corresponding to the processes that are to be simulated. At energy scales in which full field theoretic computations become relevant, particle number conservation does not hold \cite{Peskin, Maggiore}. The objective of a simulation is to track the evolution of a physical state of the theory so that the probabilities (or cross sections in the case of particle physics problems) of finding the system in various possible final states can be computed. The initial and final states usually have a finite number of quanta (particles) in them, but for a field-theoretic computation, the number of particles that may appear in the intermediate states is unconstrained. When the field theory is such that a perturbative approach is feasible, it is possible to order the contributions to the cross section of interest in terms of the number of particles in the intermediate states. The language of Feynman diagrams provides a pictorial and intuitive way of performing this ordering \cite{Feynman_diagram, Peskin, Maggiore}. However, it may be noted here that there are field theories like Quantum Chromodynamics \cite{QCD_resource_letter, Peskin, Maggiore} that do not admit a perturbative approach.  Quantum simulation itself proves to be the only efficient scheme to be used for such a study in the non-perturbative regime even though such simulations on a large scale remain to be demonstrated ~\cite{light_front}.  

    Even if the number of particles in the intermediate states is constrained, the momenta carried by the particles in the intermediate states are again unconstrained. This is because, in the usual quantization procedure for fields that splits space-time into a collection of "equal-time" slices, particles can have both positive and negative values for their momenta even if the available states are quantized in terms of their energies and momenta. Imposing energy-momentum conservation on the overall multi-particle state does not therefore constrain the number of momentum states that the particles in the intermediate states can occupy. There are no systematic means of truncating the set of states to be considered so as to allow mapping of such a truncated Hilbert space of states to the finite-dimensional Hilbert space of a register of qubits. 

    In addition to the conceptual issue of suitably truncating the Hilbert space dimension, the challenge of simulating quantum field theories is further accentuated by the limited number of qubits and coherence times of present-day NISQ devices. In fact, even very simple simulations require thousands of qubits which is not quite within the reach of present-day devices \cite{Q_sim_HEP_review, QC4HEP_grp_review}. In spite of these challenges, considerable progress has been made already in this direction for most of these many-body simulations \cite{Nachman, Qsim_not_LF_ref, QC_lattice_gauge_1, QC_lattice_gauge_2, condensedmatter_lattice_gauge, Qchem_VQE, condensedmatter_VQE}. The methods employed range from a direct adaptation of classical lattice gauge theory techniques ~\cite{Rothe, lattice_QCD} to those using the variational quantum eigensolver ~\cite{Peruzzo_VQE, Tilly_VQE_review}. Since fields are extended objects occupying all of space and time, approaches like lattice gauge theory that uses a discrete version of space-time typically require a very large number of qubits, each representing a point in space, and are beyond the capabilities of NISQ devices. We therefore choose to use the momentum-space representation of the fields with a cutoff (which is also given in terms of momentum states) as discussed above, even though all-to-all connectivity may be required between the qubits for implementing the computation using the momentum basis.

    In this paper, we focus on a particular formulation of quantum field theories called \textit{light front quantization} \cite{Pauli_Brodsky_1993, Pauli_Brodsky_2001, DLCQ_1, DLCQ_2, review_DLCQ, LF_addl_1, LF_addl_2, LF_book_chapter, LF_addl_3, LF_addl_4} that allow for a systematic truncation of the Hilbert space of the quantum fields. A reference frame traveling in a particular direction at the speed of light is chosen for quantizing the fields. Here, in contrast to the equal-time case, the dynamical variables refer to physical conditions on a light front, which is a three-dimensional surface in space-time formed by a plane wave front advancing with the velocity of light~\cite{Dirac}.  In 1 space and 1 time dimension this ensures that all massive particles are necessarily moving in one direction. The problem of particles in the intermediate states having both positive and negative momentum is avoided by this choice. As noted in \cite{Weinberg}, \cite{Susskind}, going to a frame like the light-front frame in which the total momentum of the system approaches infinity, allows one to revive the methods of "old-fashioned perturbation theory" that does not lump together intermediate states with different numbers of particles and antiparticles together. Even in the non-perturbative case, this allows for ordering and considering intermediate states with different particle numbers separately. 
    
    A single parameter called the \textit{harmonic resolution} can be defined in the light-front formulation which allows for truncating the Hilbert space dimension of the theory based on its value. It must be emphasized that the Hilbert space dimension of the full theory, and indeed, even the dimension of the sub-space relevant for a particular computation does not change whether equal-time quantization or light-front quantization is adopted. The latter only provides a means of systematically truncating the dimension based on the value of a well-defined parameter and it also allows for investigating the errors introduced by such truncation in a methodical manner. Such a truncation of Hilbert space and estimation of errors made at each level of truncation is very much in the same spirit as the standard methods used in simulating quantum chemistry problems using quantum computers \cite{Wilson}.

    Light-front quantized field theories in 1+1 dimensions have been simulated on NISQ devices ~\cite{BLFQ_ref_10, BLFQ_ref_4, BLFQ_ref_7, light_front, BLFQ_ref_9, BLFQ_ref_5, BLFQ_ref_11} and otherwise \cite{BLFQ_ref_1, BLFQ_ref_2, BLFQ_ref_3, BLFQ_ref_6, BLFQ_ref_12, BLFQ_ref_13, BLFQ_ref_8, BLFQ_ref_14}, earlier. Previous works on NISQ devices focused mostly on theories that involved only Fermionic fields. In~\cite{light_front} the light front quantization scheme and the quantum algorithm for simulation for a theory involving a Boson and a Fermion, anti-Fermion pair is developed. We study the time evolution and the dynamics of this $(1+1)$ dimensional Yukawa model. The first forays into quantum simulation of Quantum Chromodynamics (QCD) deal with processes involving quarks that are bound inside nucleons. At "low-enough" energies within the QCD energy scale, it is possible to consider a theory of only the Fermionic components of the nucleons and indeed there are a few similarities between the treatment of the nucleons bound inside the nucleus and electrons bound inside atoms that are at the center of quantum chemistry computations. At slightly higher energy scales, particles like Pions become relevant for the QCD calculations. The $(1+1)$ dimensional Yukawa model considered in~~\cite{light_front} aims to capture the interesting physics at this energy scale. Our focus here is more on the technical aspects of implementing such a computation on a NISQ device and on clarifying the limitations as well as the way forward. For the simulation of the field theory, we require an implementation of both Fermionic and Bosonic operators on the NISQ device. The Bosonic field operators are particularly challenging to implement on such devices and our implementation suggests one path forward. The circuit construction and execution were done using Qiskit (version $0.39.5$) \cite{Qiskit} – IBM’s quantum computing SDK, using IBM Quantum Experience cloud-based access \cite{IBMQ}. 
	
	\section{Light-front formulation of a (1+1)D field Hamiltonian}
    \label{sec1}

\vspace{2 mm}
    The model we consider has a Bosonic field $\phi$ with quanta of mass $m_B$ coupled to a Fermion field $\psi$ with particles and anti-particles both with mass $m_F$. The fields are coupled through a Yukawa type term, $\lambda \phi \bar{\psi} \psi$. The theory is described by the Lagrangian density, \cite{Pauli_Brodsky_1993, Pauli_Brodsky_2001}
	\begin{eqnarray}
		\label{L}
			\mathcal{L} & =  & \frac{1}{2}\partial^\mu\phi\partial_\mu\phi - \frac{1}{2}m_B^2\phi^2 \nonumber \\ 
			& & \quad + \frac{i}{2}\bar{\psi}\gamma^\mu\partial_\mu\psi - \frac{i}{2}\partial_\mu\bar{\psi}\gamma^\mu\psi - m_F\bar{\psi}\psi\nonumber \\ 
            && \qquad - \lambda \phi \bar{\psi}\psi,
	\end{eqnarray} 
    where $\lambda$ is the bare coupling strength. Light-front quantization of the theory is described in detail in~\cite{Pauli_Brodsky_1993, Pauli_Brodsky_2001,light_front} and we summarise this discussion below for completeness. 
    
    In 1+1 dimensions, the transformation to light-front coordinates is given by
    \[x^+ = x^0+x^1 \quad {\rm and} \quad x^- = x^0-x^1,\]
    where $x^0$ and $x^1$ are time and position coordinates in the lab frame while $x^+$ and $x^-$ are the light-front time and light-front position coordinates respectively. The change of frame modifies the energy-momentum conservation relation in the lab frame, $k_\mu k^\mu = m^2$ to 
    \begin{equation}
        \label{energymomentum}
        k^+k^- = 4k_-k_+ = m^2.
    \end{equation}
    This corresponds a change in the components of the metric tensor from $g^{00} = -g^{11} =1$, $g^{01}=g^{10}=0$ in the lab frame to $g^{++}=g^{--}=0$, $g^{+-} = g^{-+} = 2$ in the light-front coordinates. In the rotated frame, although the meaning of energy and momentum is lost, one can still associate a meaning of light-cone momentum and light-cone energy of a single particle to $k^+ \equiv k_0 + k^1$ and $k^- \equiv k^0 - k^1$, respectively. In the lab frame, for a given momentum $k^1$, the allowed values of energy are $k^0 = \pm \sqrt{m^2 + (k^1)^2}$, with the two values associated with the particle and antiparticle respectively. In the light-front frame, we have already noted that the momentum $k^+$ has only one sign since all particles should appear to move in one direction only. From ~\eqref{energymomentum}, we see that the light-front energy $k^- = m^2/k^+$ also is exclusively all positive or all negative for particles. For anti-particles, the signs of both $k^+$ and $k^-$ will be reversed.  

    It is customary to discretize the light-cone momentum by invoking quantization within a box of length $L$ as~\cite{Pauli_Brodsky_1993},
    \begin{equation}
    \label{kplus}
    k^{n+} = \frac{2 \pi}{L} n, \quad n=1,\ldots, \Lambda,
    \end{equation}
    which is a second-quantized formulation of the theory, known as the discretized light-cone quantization (DLCQ).
    
\begin{figure}[!htb]
			\includegraphics[width=0.99\textwidth]{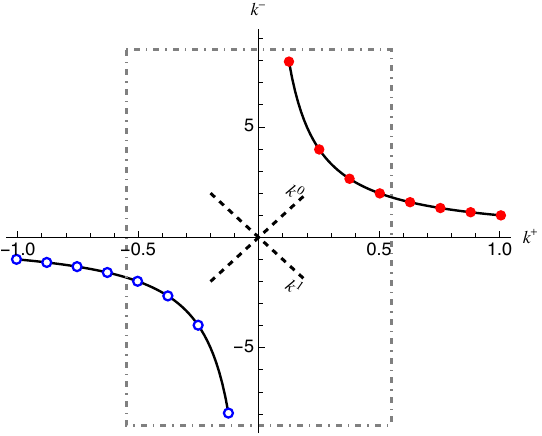} 
		\caption{The discretized field modes for particles (solid, red circles) and for anti-particles (open, blue circles) are shown in the $k^+-k^-$ plane. The dashed lines show the $k^0$ and $k^1$ axes respectively as seen in the light-front frame. The box demarcated with dot-dashed grey lines shows the effect of imposing a cutoff $\Lambda$ on the harmonic resolution (shown here for $\Lambda=4$). The states lying within the box for both particles and anti-particles are the only ones that need to be considered, provided the initial and final physical (multi-particle) states that we are interested in have light-front momentum less than the cutoff. Note that the number of states at low values of $k^+$ is regulated by the mass of the particles which is assumed to be non-zero. A renormalization of the reference state of the particles, or the Fock space vacuum, can shift the states within the entire dot-dashed box upwards (See Appendix~\ref{appA} for details) and make the light-cone energies of all the states within the box positive.} \label{fig1}
	\end{figure}

    The space of states for particles and anti-particles is shown diagrammatically in Fig.~\ref{fig1}. Choosing the cutoff $\Lambda$ limits the number of states that have to be considered in any simulation of allowed processes in the theory. The cutoff $\Lambda$ sets the harmonic resolution. At lower values of $k^+$, the number of states is regulated by assuming that all the fields are massive. While the cutoff $\Lambda$, as well as the box-length $L$, can be imposed in the equal-time representation of the field also, as mentioned earlier, the momentum of component particles having both signs means that energy-momentum conservation for physical initial and final states that we are interested in does not limit the number of intermediate states that have to be included in the simulation. It is also worth noting that by employing a suitable renormalization of the masses of the particles/anti-particles, one can make the energies of all states considered up to the cutoff $\Lambda$ of the harmonic resolution positive, further simplifying the mapping of the field modes on to qubits. 

    There are two independent free fields $\phi_{\rm f}$ and $\psi^{(+)}_{\rm f}$ in the non-interacting theory which can be expanded into plane waves in the light-front frame on the interval $(-L, L)$ as.
	\begin{subequations}
		\label{field_solns}
		\begin{equation}
			\phi_{\rm f}(x^\mu) = \sum_{\substack{n=1 \\ \mu = \pm}}^{\Lambda} \frac{1}{\sqrt{4\pi n}}\big( a_n e^{-ik_\mu^{n}x^\mu} + a_n^\dagger e^{+ik_\mu^{n}x^\mu} \big),
		\end{equation}
		and
		\begin{equation}
			\psi_{\rm f}^{(+)}(x^\mu) = \frac{u}{2L}\sum_{\substack{n=1 \\ \mu = \pm}}^{\Lambda} \big( b_n e^{-ip_\mu^{n}x^\mu} + d_n^\dagger e^{+ip_\mu^{n}x^\mu} \big).
		\end{equation}
	\end{subequations} 
	In the equation above, 
    \[ \psi^{(\pm)} = \Gamma^{(\pm)} \psi = \frac{1}{4} \gamma^{\mp} \gamma^{\pm} \psi,\] 
    where $\gamma^{\pm} = \gamma^0 \pm \gamma^1$ are the transformed $\gamma$-matrices in the light-front frame with $\gamma^0 = \sigma_3$ and $\gamma^1 = i \sigma_2$. Note that the spinor $u$ appearing in the plane wave expansion of $\psi^{(+)}$ is independent of the momenta and is normalized to unity so that, $u = (1, 1)^{\rm T}/\sqrt{2}$. The light cone momenta $k_+^n$ and $p_+^n$ appearing in ~\eqref{field_solns} are given by ~\eqref{kplus}, and corresponding to these, we have the discretized light-cone energies for the Bosons and Fermions,
	\begin{equation}
		\label{k-}
		k_-^n = \frac{m_B^2}{k_n^+}, \quad \text{ and } \quad  p_-^n = \frac{m_F^2}{p_n^+}.
	\end{equation} 
    The Bosonic and Fermionic creation and annihilation operators appearing in the expressions for the free fields satisfy the usual commutation and anti-commutation relations respectively:
     \[ \{b_n, b_m^\dagger\} = \delta_{n,m} \quad \text{and} \quad  \{d_n, d_m^\dagger\} = \delta_{n,m},\] 
     for the Fermions and anti-Fermions, and 
     \[ [a_n, a_m^\dagger] = \delta_{n,m}, \]
     for the Bosons, with all other commutators being zero. 

    In the lab frame, the complete set of commuting operators (CSCO) whose eigenvalues label the physical states in the theory are the total charge, $Q$, total momentum, $P$ (or $P^1$), and total energy (Hamiltonian), $E$ (or $P^0)$. The operator corresponding to the total charge, $Q$, remains unchanged under the transformation to the light-front frame. However, $P$ and $E$ operators are replaced by $P^+$ and $P^-$ corresponding to the dynamical variables $k^+$ and $k^-$, and related as $P^+ = P^0 + P^1$ and $P^- = P^0 - P^1$. In line with ~\eqref{kplus}, we can introduce a pair of operators $K$ and $H$ as,
    \[ P^+ = \frac{2 \pi}{L} K, \quad \text{and} \quad P^- = \frac{L}{2\pi} H,\]
    where $K$ is identified as the modified momentum operator and $H$ is identified as the modified energy operator. Note that we can also define the square of the invariant mass, $M^2$, in space-time quantization in terms of the total momentum and energy operators, $P^0$ and $P^1$, as $M^2 = (P^0)^2 - (P^1)^2$. The corresponding operator in light-cone quantization is hence $M^2 = P^+P^- = KH$ with the operator $M^2$ having no dependence on the length scale, $L$, used for discretizing the momenta. The operator $K$ is dimensionless whereas $H$ has the dimension of a mass squared. In terms of the Bosonic and Fermionic creation annihilation operators, we obtain, 
    \begin{equation}
		\label{Q}
		Q = \sum_n (b_n^\dagger b_n - d_n^\dagger d_n),
	\end{equation}
	and
	\begin{equation}
		\label{K}
		K = \sum_n n(a_n^\dagger a_n + b_n^\dagger b_n + d_n^\dagger d_n),
	\end{equation}
    where $Q$ can be viewed as the Baryon number.
    We see that both operators are diagonal in the number basis. The operator $H$ is not diagonal and it is made up of four terms as
    \begin{equation}
        \label{eq:Hamil1}
        H = H_M + H_V + H_S + H_F.
    \end{equation}
    The four terms in $H$ are worked out in detail in \cite{Pauli_Brodsky_1993} and they are also given in Appendix~\ref{appA} for easy reference. The mass term $H_M$ in the Hamiltonian, which is also diagonal in the number basis, contains the renormalized masses that make the energies of all the states that are being considered within the cutoff $\Lambda$ positive, as mentioned earlier. 
	
	\section{Simulation of the (1+1)D field:}
 \vspace{2 mm}

    With the cutoff on the Harmonic resolution, $\Lambda$, in place, for each species of particle, we have a finite number of states to work with. As basis vectors in this Hilbert space, we chose Fock states specified in the form,
    \[ |n_1, \ldots, n_{N_F}; n_1', \ldots, n_{N_A}'; p_1, \ldots, p_{N_B} \rangle, \]
    where $n_i$, $n_i'$, and $p_i$ denote the Fermonic, anti-Fermionic, and Bosonic states and their respective occupancies. The cutoffs $N_F$, $N_A$, and $N_B$ denote the number of states of each species that is considered. For Fermions and anti-Fermions, the occupancies $n_i$ and $n_i'$ can either be 0 or 1 only. For Bosons, however, each mode can be multiply occupied with the multiplicity referred to here as the number of modals~\cite{Ollitrault}. The global cutoff $\Lambda$ indirectly limits the number of modals to be considered as well. The multiplicity (number of modals) of each mode of the Boson field is denoted as $m_1, \ldots, m_{N_B}$.

	Now that the choice of basis to span the Hilbert space of states of the field theory has been fixed, we can estimate the qubit resources required to map these states. The number of qubits required to map $m$ modals will be $\lceil \log_2 (m+1) \rceil$. Assigning each mode to a qubit, the Fermionic and anti-Fermionic states would require $N_{F}$ and $N_{A}$ qubits respectively. The Bosonic field with $N_B$ modes, each having multiplicity $m_i$ $(i \in 1, \ldots, N_B)$, will require $\sum_1^{N_B} \lceil \log_2 (m_i+1) \rceil$ qubits. 
	Hence, generic states of the three fields can be mapped into corresponding states of a register with $N_F + N_A + \sum_1^{N_B} \lceil \log_2 (m_i+1) \rceil$ qubits in total. For the case where all the Bosonic modes have the same multiplicity $m$, this value will modify as $N_F + N_A + \lceil \log_2 (m+1) \rceil N_B$.

    The number of available qubits in a NISQ device, in practice, limits the maximum values of $N_F$, $N_A$, $N_B$, and $m$ that we can consider. This, in turn, sets the practical limit on the maximum value of $K$ from Eq.~\eqref{K} that we can consider, irrespective of the theoretical cutoff $\Lambda$. However, as pointed out in~\cite{light_front}, there are problems of interest in which only low values of $K$ appear, and so for such problems, the cutoff imposed by the number of available qubits does not lead to significant errors.
	
	\subsection{Particle operators and their mapping}
	
	The computations in the following have been carried out on the simulators and quantum processors provided by IBM cloud-based access, using IBM Qiskit \cite{Qiskit}. We have identified the qubit requirements for mapping Fock states of the theory and as the next step, the creation and annihilation operators which can act on the Fock states have to be implemented on the qubit register. Since Fermions and anti-Fermions both follow Fermi-Dirac statistics, we can use the same operators for both of them. Here, Fermionic operators  (FermionicOp) defined in the Qiskit Nature module~\cite{qkit_nature} under the second quantized operators (qiskit\_nature.second\_q.operators.FermionicOp) have been used for these particles. 
	These are single-qubit operators, $b$ and $b^\dagger$ respectively, whose action on single-qubit states satisfy the Pauli-exclusion principle. Identical operators,  $d$ and $d^\dagger$, serve as creation and annihilation operators for anti-Fermions. 
	Since the particles that the qubits are standing in for have spin also, the Fermionic creation and annihilation operators need to be mapped onto spin operators as well. Out of several mapping schemes available for this purpose, the Jordan-Wigner mapping \cite{JW} was chosen here, which maps each Fermionic operator onto a qubit \cite{qkit_nature}. \\
	
	For the Bosons, since each mode can be multiply occupied, the operators also have to incorporate this multiplicity. For a Bosonic mode the creation and annihilation operators ($a^\dagger$ and $a$ respectively) are of the form \cite{Zettili}, 
	\[ a^\dagger|l\rangle = \sqrt{l+1}|l+1\rangle \;  \text{and} \; a|l\rangle = \sqrt{l}|l-1\rangle, \; l=0,1,\ldots\]
	
	However, for practical purposes, since infinite dimensional representations cannot be used in simulations, they are truncated so that $a^\dagger|m\rangle = 0$ \cite{bosons_arxiv}, where $m$ is referred to as the number of \textit{modals} of the given field mode. Corresponding to this mode, the Hilbert space of states of the Bosonic field therefore contains $m+1$ states which may be labelled as $|0\rangle, |1\rangle, \ldots, |m\rangle$ for each mode. We do not carry the mode label in the following since the identity of the modes we are considering is usually obvious from the context.  A single mode of the Bosonic field with $m$ modals requires  $t = \lceil \log_2 (m+1) \rceil$ qubits for representing its basis states. The mapping is easily obtained as the binary representation of the integers $0, 1, \ldots, m$. We chose $m= 2^t-1$ so that the $m+1$ basis states can be mapped onto the states of $t$ qubits. For instance, if $m=7$ we have $|0\rangle \leftrightarrow |000\rangle, |1\rangle \leftrightarrow |001\rangle, \ldots, |7\rangle \leftrightarrow |111\rangle$. The operators corresponding to $a$ and $a^\dagger$ for each mode are constructed out of the four single qubit operators $\sigma_+, \sigma_-, I_+, \text{ and } I_-$  constructed from the Pauli matrices, and defined in \cite{bosons_arxiv} as
	\begin{eqnarray}
    \label{boson_sigma_basis}
	    \sigma_+ &  = & (\sigma_x + i\sigma_y)/2= |0\rangle \langle 1|, \nonumber \\  
        \sigma_- &  = &  (\sigma_x - i\sigma_y)/2= |1\rangle \langle 0|, \nonumber \\
	    I_+ &  = & (I + \sigma_z)/2 = |0\rangle \langle 0|,  \nonumber \\
        I_- &   = &  (I - \sigma_z)/2 = |1\rangle\langle1|.
   \end{eqnarray}
   For instance, in terms of these operators, the equations $a^\dagger|0\rangle = \sqrt{1} |1\rangle$ and $a^\dagger|3\rangle = \sqrt{4}|4\rangle$, for the $m=7$ case, turns into $\sqrt{1} I_+ \otimes I_+ \otimes \sigma_-|000\rangle = \sqrt{1} |001\rangle$ and $\sqrt{4} \sigma_-\otimes \sigma_+ \otimes\sigma_+|011\rangle = \sqrt{4}|100\rangle$ respectively. 

	While there is no pattern that is immediately obvious in mapping the Bosonic creation and annihilation operators into strings of single qubit operators from the set $\{I_\pm, \sigma_\pm\}$, we propose the following way of representation for the mapping: 
	\begin{equation}
		\label{bosonic_op}
        a^\dagger = \sum_{p = 0}^n \Big[\sum_{j_p}\sqrt{j_p} \, {\mathcal I}_{j_p^{\text{bin}}}^{\otimes p} \Big] \otimes \sigma_ - \otimes \sigma_+^{\otimes(n-p)}.
	\end{equation}
	The notation used in the expression above is quite condensed and requires a bit of explanation. The upper limit of the sum is one less than the number of qubits required to represent the modals, $n = \lceil \log_2 (m+1) \rceil - 1 = t-1$ and we also define $q = n - p$. For each value of $p$, the integers $j_p$ varies in steps of $2^{q+1}$ in the range $2^q, ..., m$. For example, if $m=7$ with $t=3$, $n=2$ and $p=1$ ($q=1$), then $j_p$ runs in the range $2$ to $7$, with a step size of $4$. This means that the only values possible for $j_p$ in this case are $2$ and $6$. The operators ${\mathcal I}_{j_p^{\text{bin}}}^{\otimes p}$ represents a tensor product of $p$ copies of either $I_+$ or $I_-$. The operator ${\mathcal I}$ is indexed by the digits appearing in the binary form of the integer $j_p$ (with the length of this bitstring being equal to $n$) with the mapping $0 \rightarrow +$ and $1 \rightarrow -$. For each value of $p$, the binary expansion of $j_p$ is always assumed to have as many digits as required to hold its largest value. In other words, the digits appearing in $p$ when written in binary form determine whether $I_+$ or $I_-$ is used in each position in the expression for ${\mathcal I}$. If the binary expansion of $j_p$ has more digits than $p$ itself, then the first $p$ digits are used. 
 
    In our example with $m=7$ and $p=1$, we have $j_1$ equal to either $010$ or $110$ in binary. We have taken three digits in the expansions of $2$ and $6$ since the binary representation of $6$ requires all three digits. Since $p=1$, only the first digit in each of these binary sequences (0 and 1 respectively) are used for constructing ${\mathcal I}$ and we have for this case, 
    \[ \sum_{j_1}\sqrt{j_1} \, {\mathcal I}_{j_1^{\text{bin}}}^{\otimes p}  = \sqrt{2}I_+ + \sqrt{6}I_-.\]
    This means that $p=1$ contributes the two terms $\sqrt{2}I_+ \otimes \sigma_-  \otimes \sigma_+$ and $ \sqrt{6}I_- \otimes \sigma_- \otimes \sigma_+$ to the mapping of $a^\dagger$ for $m=7$. When $p=0$ ($q=2)$, $j_p$ ranges from $4$ to $7$ in steps of $2^{2+1}=8 > m$, which means that the only allowed value of $j_0$ is $4$ which is equal to $100$ in binary. However here the binary expansion of $4$ is moot since ${\mathcal I}$ is a string of length zero. Hence, $p=4$ contributes only a single term, namely $\sqrt{4} \sigma_- \otimes \sigma_+ \otimes \sigma_+$ to $a^\dagger$. Finally when $p=2$ we have $j_2$ in the range $2^0=1$ to $m=7$ in steps of $2^1 =2$. The possible values are $j_2 = 1,3,5,7$ which, in binary are $001$, $011$, $101$, and $111$ respectively, of which we have to take the first two digits, namely, $00$, $01$, $10$ and $11$. So we have  
    \begin{eqnarray*} 
    \sum_{j_2}\sqrt{j_2} \, {\mathcal I}_{j_2^{\text{bin}}}^{\otimes p}  & = &  \sqrt{1}I_+\otimes I_+ + \sqrt{3}I_+ \otimes I_-  \\ 
    && \qquad + \, \sqrt{5}I_- \otimes I_+ + \sqrt{7} I_-\otimes I_-.
    \end{eqnarray*}
    
	Using the results above, we obtain for $m=7$,
    \begin{eqnarray*}
			a^\dagger & = & \sqrt{1} I_+ \otimes I_+ \otimes \sigma_- + \sqrt{2} I_+ \otimes \sigma_- \otimes \sigma_+ \\
			& & \quad + \sqrt{3} I_+ \otimes I_- \otimes \sigma_- + \sqrt{4} \sigma_- \otimes \sigma_+ \otimes \sigma_+ \\
			& & \quad + \sqrt{5} I_- \otimes I_+ \otimes \sigma_- + \sqrt{6} I_- \otimes \sigma_- \otimes \sigma_+ \\
			& & \quad + \sqrt{7} I_- \otimes I_- \otimes \sigma_-. 
	\end{eqnarray*}

    Alternatively, one can also realize that ${\mathcal I}_{j_p^{\text{bin}}}^{\otimes p}$ denotes the tensor product of $p$ number of $I_+$ or $I_-$ operators such that terms with all possible combinations of these operators ($2^p$ terms) are included in the summation. This is similar to creating binary strings of a given length $p$. For instance, if $p = 2$, there will be $2^p = 4$ possible binary strings - $00, 01, 10, \text{ and } 11$, as well as $4$ permissible values for $j_p$. Considering $I_+ \equiv 0$ and $I_- \equiv 1$, the four ${\mathcal I}_{j_p^{\text{bin}}}^{\otimes p}$ terms here for $p = 2$ will be $I_+I_+$, $I_+I_-$, $I_-I_+,$ and $I_-I_-$, in the ascending order according to this convention of comparing with the binary values, and these will have the associated $j_p$ values corresponding to the these terms, also in the ascending way, as $1, 2, 3,$ and $4$. For a particular value of $p$, the value of $q$ will remain fixed according to the relation $p+q = n$,  while the value of $j_p$ progressively changes for each ${\mathcal I}_{j_p^{\text{bin}}}^{\otimes p}$ term in the range, as defined above. 
    
	The Bosonic annihilation operator is the adjoint of this creation operator. One can easily verify that these creation and annihilation operators satisfy the required commutation relations. 
    
	\subsection{Hamiltonian time evolution:}
	
	With the Fock space and the necessary second quantized operators for the different particles defined, the Hamiltonian can now be mapped onto the qubits. Since both Bosonic and Fermionic operators are mapped onto strings of Pauli operators, individual terms in the Hamiltonian will also be strings of Pauli operators. The Hamiltonian itself is therefore a sum of Pauli operator strings making it an object that belongs to the Qiskit PauliSumOp class. The infinitesimal changes generated by operators of this class on the quantum states of a qubit register can be conveniently approximated without requiring the resource-prohibitive step of conversion of the operator into the matrix form. In our case, we are interested in the exact time evolution generated by the Hamiltonian that is obtained by computing the exponential of the time-independent Hamiltonian given in Appendix \ref{appA} to obtain the unitary time evolution operator $U(t) = \exp [ -i H t / \hbar ]$.
	
	The individual Pauli operator strings in the Hamiltonian do not commute with one another. So to find $U(t)$ we use the Suzuki-Trotter approximation~\cite{Trotter, Suzuki, Suzuki_decomp} which is utilized to perform the Trotter expansion procedure to $k^{\text{th}}$ order \cite{qkit_opflow}. The time evolution from $0$ to $t$ is broken up into $n_{\text{T}}$ steps and for $k^{\text{th}}$-order Suzuki-Trotter formula, the error incurred with respect to the exact evolution is $\sim n_{\text{T}}^{-2k}$ \cite{Preskill}. However, higher-order approximations increase the quantum circuit depth leading to increased decoherence effects, and an effective trade-off between the two needs to be found. Further discussion on the characterization of Trotter errors can be found in \cite{trot_err_1}\cite{trot_err_2}, and on Trotter error mitigation in \cite{trot_err_mitigate_1}. Note that the Bosonic creation and annihilation operators consist of sums of products of $\sigma_+$, $\sigma_-$, $I_+$, and $I_-$, that, in turn, are given in terms of the Pauli operators by Eq.~\eqref{boson_sigma_basis}. The Trotterization steps available in Qiskit (the PauliTrotterEvolution class) would create products of exponents of terms containing the Pauli operators after using Eq.~\eqref{boson_sigma_basis} for expanding $\sigma_+$, $\sigma_-$, $I_+$, and $I_-$. Alternate schemes for trotterization can be utilized instead, such as those that can keep the $\sigma_+$, $\sigma_-$, $I_+$, and $I_-$ operators together, so that no term in the bosonic operator is split further. This can aid in making the trotterization approximation easier and less prone to errors that lead to violations of the conservation laws. In the simulations executed here, further reduction was done to the Pauli string terms in the Hamiltonian, in order to combine similar operator strings coming from the various terms in the Hamiltonian. This also ensures that no additional cost is inadvertently added to the circuit due to duplication of terms. The trotterized operator was then created as an object belonging to the Qiskit PauliTrotterEvolution class, which transforms the exponential of the Hamiltonian, which is a sum of Pauli terms, to products of exponents of these Pauli terms. This parameterized operator, with the evolution time as the parameter that can be varied, was then mapped to the circuit, in order to execute the trotterized evolution of the input state. Before execution. the circuit was also transpiled with the highest optimization\_level value of $3$, in order to execute further possible decomposition and optimization. The last step of the simulation is to read out the qubit register(s) to find the probabilities for various final field configurations. 
   	
	\section{Results:}
	
	The focus of our simulations is to verify that the steps outlined above can be implemented in principle on a NISQ device. The quantum computational resources available are clearly not adequate to perform a satisfactory simulation of a realistic process described by the Hamiltonian we consider. So we have chosen to mention realistic parameter values applicable to the theory only very briefly just to set the scale of the problem and then chose to use arbitrary units in which the parameters take on simple numerical values. We believe this choice places the focus on the algorithm and technique behind the implementation of the simulation rather than being distracted by semi-realistic concerns. However, we note here that this approach has the drawback that possible comparisons with experimentally observed values are limited to comparing qualitative trends only.  
 
    The field theory we consider is a model for the well-known and extensively studied example of Proton-Pion scattering. Since both the $\pi^+$ and $\pi^-$ couple identically with the proton, the Pion field in our study was generically taken to be a single field denoted as $\pi^\pm$ (the charged-pion). No further distinction between the two kinds of charged Pions was made. The charges do not play a significant role in our toy model since no coupling with the electromagnetic field is included in it. The bare Fermion and Boson masses, $m_F$ and $m_B$, were taken to be equal to their physical masses, $\tilde{m}_F$ and $\tilde{m}_B$, respectively, as in \cite{Pauli_Brodsky_2001}, and we chose to work in pion mass ($m_\pi = 139.57$ MeV) units so that $\tilde{m}_F = 6.7$ and $\tilde{m}_B = 1.0$. Note that technically to consider only positive energy states, a mass renormalization as mentioned in Sec.~\ref{sec1} has to be applied. However, since we are working in arbitrary units, this additional, straightforward correction to the masses or to the coupling constant \cite{Pauli_Brodsky_2001} \cite{Brooks1984} is avoided here for the sake of clarity.  
	
    The remaining parameters to be fixed are the bare coupling constant, $\lambda$, between the Fermionic and Bosonic fields, the maximum number of modes, $N_{\text{max}}$, to be considered for each particle species, assuming equal number for each, the duration of time, $\Delta t$, corresponding to each Trotter step and the number of Trotter steps, $n_{\text{T}}$. Also to be determined are the number of modals for each Bosonic mode and the cutoff $\Lambda$ for the total momentum. In the following, we explore how the simulation results vary both in terms of accuracy and repeatability when $\lambda$, $N_{\text{max}}$, $\Delta t$ and $n_{\text{T}}$ are varied while keeping the number of modals and the cutoff, $\Lambda$, set equal to $3$ and $2048$ respectively. The cutoff, $\Lambda$ was set to a high value so as to make the results independent of it \cite{Pauli_Brodsky_2001}. In addition to this, to make the results independent of the value of the light-cone box size $L$, its value was set to $2\pi$ $m_{\pi}^{-1}$, in all cases, similar to the convention in \cite{BLFQ_ref_2}, so that the time evolution operator in light-front, $P^- = (L/2\pi) H$ becomes equivalent to the operator $H$ itself.
    
    \subsection{Analysis of Suzuki-Trotter}	
	
\begin{figure}[!htb]
		\centering
		\includegraphics[width=0.99\linewidth]{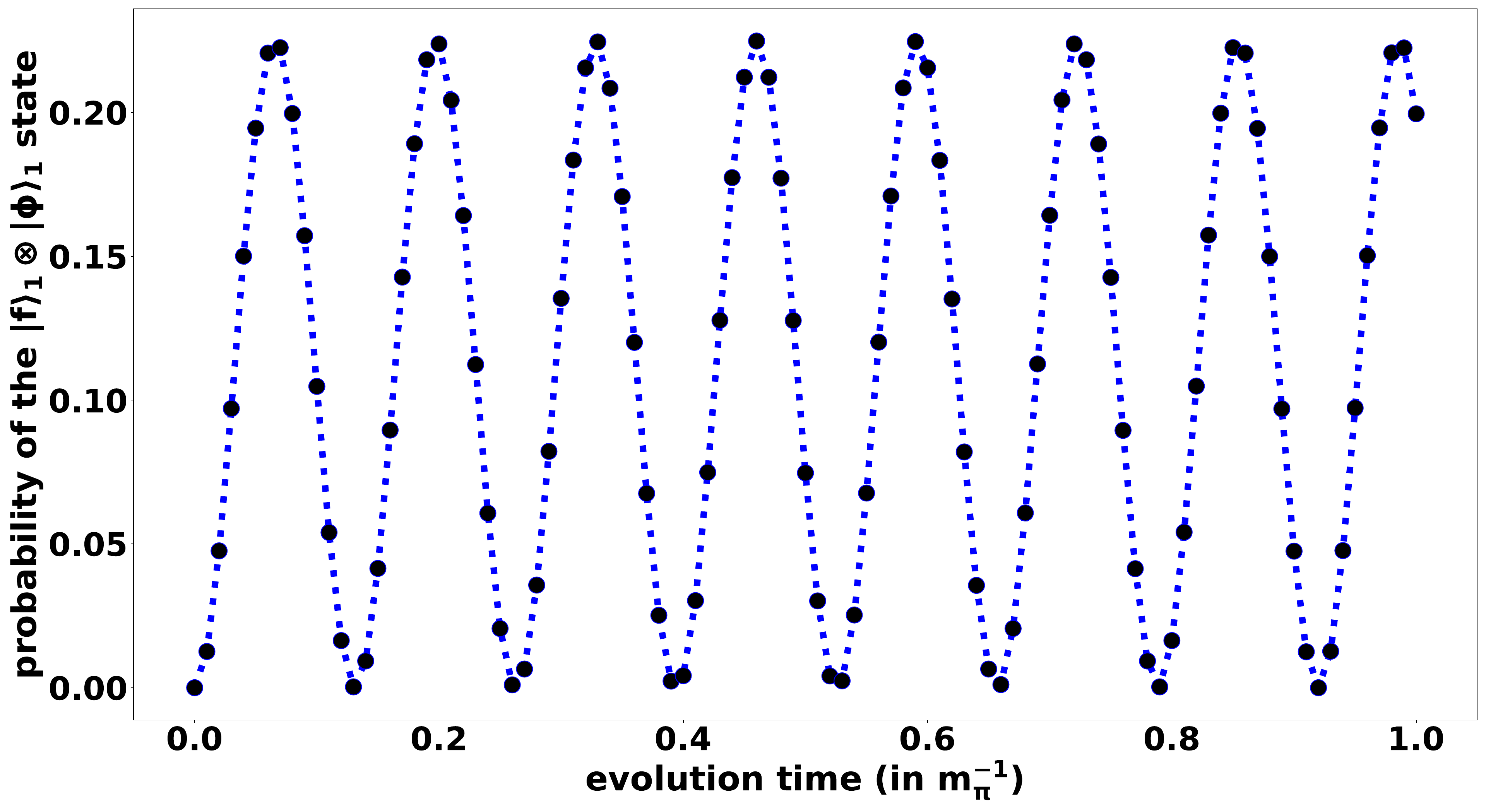} 
		\caption{The probability of obtaining the state $|f \rangle_1 \otimes |\phi\rangle_1 = |100 \: 000 \: 01 \, 00\, 00\rangle $ corresponding to exact Hamiltonian evolution is plotted as a function of time and it exhibits the oscillatory behavior between the two allowed states $|f\rangle_2$ and $|f\rangle_1 \otimes |\phi\rangle_1$, as expected. The simulations were done on a classical computer.}
		\label{exact_prob_vs_time}
	\end{figure}

    Prior to computing the time evolution of the different initial states, we analyze how well the Suzuki-Trotter approximation performs. For this, an example with a small enough state space was considered so that the exact evolution can also be performed and the results therein be used for comparison with the the results yielded by the approximation. We consider $3$ modes (i.e., $N_{\text{max}} = 3$) for each kind of particle, and $3$ Bosonic modals. Exact evolution was performed with $\lambda$ as $4.0$, over a time interval $t = 1.0$ in units of $m_{\pi}^{-1}$ with a step size of $0.01$ $m_{\pi}^{-1}$. The initial state had a Fermionic excitation in the mode $K = 2$ which is represented in the qubit register as $|010 \ 000 \ 00\, 00\, 00\rangle$. Since evolution conserves  $K$ and $Q$, the only allowed transition is of the form $|f\rangle_2 \longrightarrow |f\rangle_1 \otimes |\phi\rangle_1$, by which the Fermion undergoes a transition to  $K = 1$ while creating a Bosonic excitation with $K = 1$. This state is represented by $|100 \: 000 \: 01\, 00\, 00\rangle$, and time evolution is confined to the two-dimensional Hilbert space spanned by these two states. The transition probability, which is equal to the probability of obtaining the state  $|100 \: 000 \: 01 \, 00\, 00\rangle$, is plotted as a function of time for exact evolution in Fig.~\ref{exact_prob_vs_time}. 
	
    \begin{figure}[!htb]
		\centering
		\includegraphics[width=0.99\linewidth]{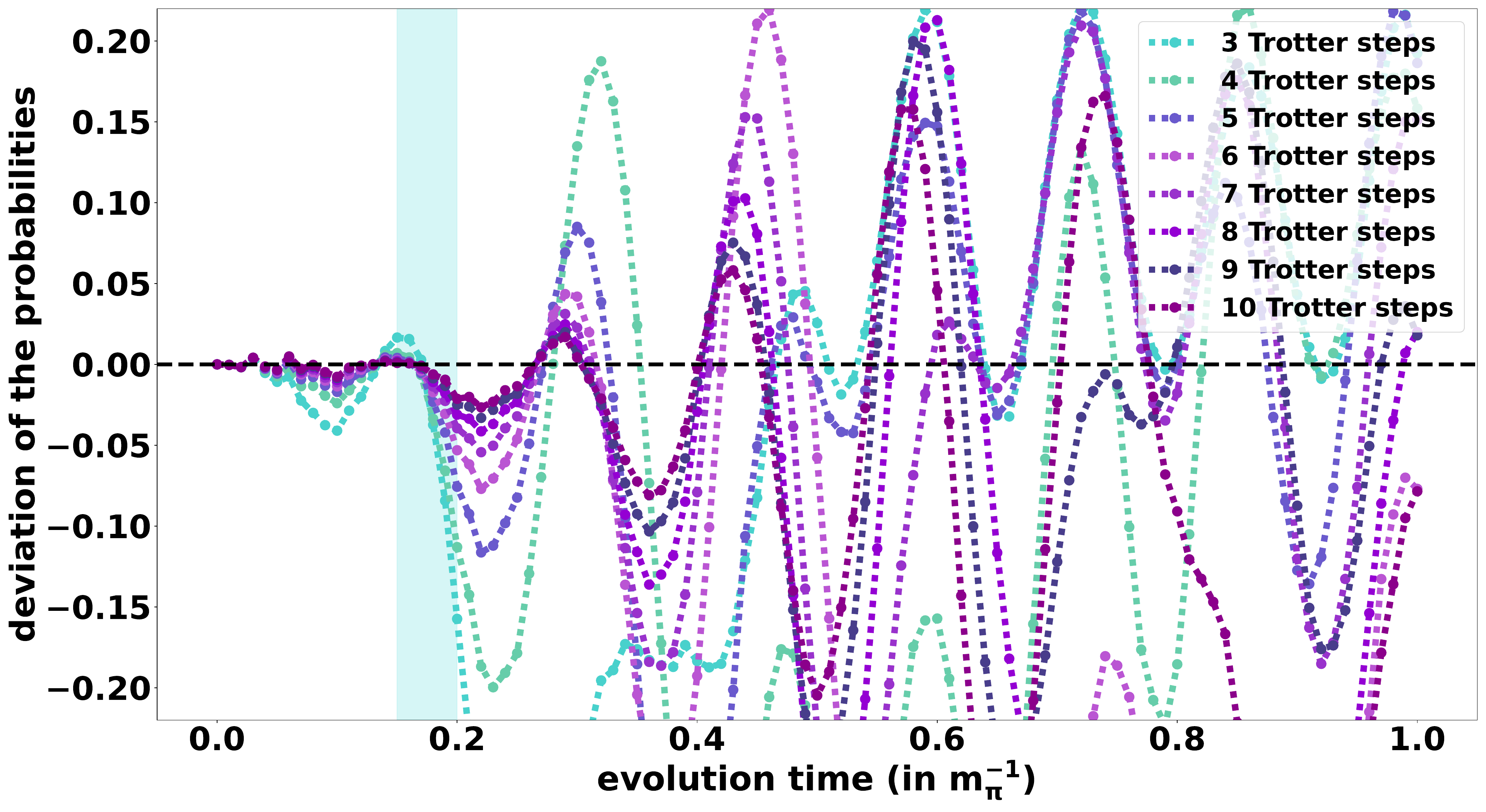} 
		\caption{The deviation of the probabilities for obtaining the state $|f \rangle_1 \otimes |\phi\rangle_1 = |100 \: 000 \: 01 \, 00\, 00\rangle$ from the exact values shown in Fig.~\ref{exact_prob_vs_time} is plotted as a function of total evolution time $t$ when the Trotter expansion with varying number of steps $(n_{\text{T}})$ ranging from $3$ to $10$ is used. The interval from $0$ to $t$ is split into $n_{\text{T}}$ steps for each value of $t$ shown. For $n_{\text{T}} = 1, 2$ the variation is huge and is not shown here.  The simulations were performed on the \textbf{ibmq\_qasm\_simulator}. Till the narrow shaded region at $t=0.2\; m_{\pi}^{-1}$,  the Trotter evolution for most values of $n_{\text{T}}$ match the exact evolution, while for larger values of the evolution time, the higher algorithmic error necessitates a further increase in the value of $n_{\text{T}}$ beyond $n_{\text{T}} = 10$.}
		\label{Trotter_exact_comparison}
	\end{figure}
 
    The cost in terms of circuit depth was the main consideration when deciding on the order of the Suzuki-Trotter approximation employed. The trotterization results for first and second orders were executed and compared for an arbitrary example, where the circuit depth for a single first-order Trotter step was found to be $14118$ for the general basis set of the \textbf{ibmq\_qasm\_simulator}, while that for a single second-order Trotter step was found to be $28223$, which is nearly double of the first-order one. So, even though the second-order approximation took roughly half the number of Trotter steps than that of the first-order one to get agreeable results, the cost of the circuit being double makes the resource requirement for both approximations identical. Therefore, in the following, only first-order Trotter approximation is employed and the focus was on optimizing the number of Trotter steps into which the evolution time was broken up.

    \begin{figure}[!htb]
		\centering
		\includegraphics[width=0.99\linewidth]{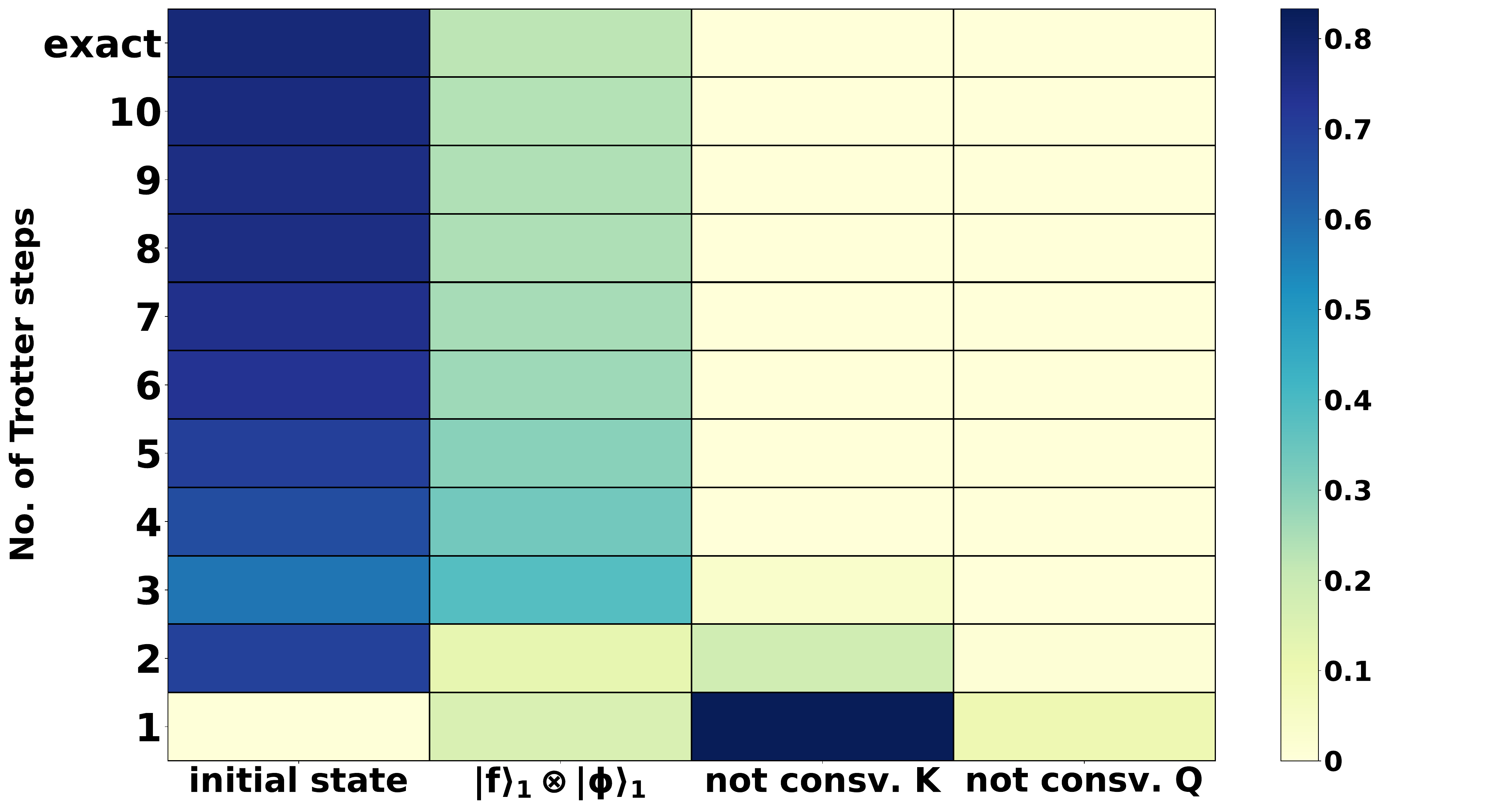} 
		\caption{Probabilities for different states obtained after evolving the initial state $|f\rangle_2 = |010 \: 000 \: 00 \, 00\, 00\rangle$ for $t=0.2\, m_{\pi}^{-1}$. The rows correspond to the number of Trotter steps into which the evolution time is broken up, with the top row representing exact evolution. The survival probability (left column), the transition probability into the state $|f\rangle_1 \otimes |\phi\rangle_1$ which conserves both $K$ and $Q$ (second column from left) as well as computed probabilities for transitions into states that do not conserve $K$ or $Q$ (last two columns) are shown. The simulations were done on the \textbf{ibmq\_qasm\_simulator}. Colors indicate probability values. We see, as observed in Fig.~\ref{Trotter_exact_comparison} also, that after about 5 trotter steps, the leakage of probability to states that do not conserve $K$ or $Q$ becomes very small. By 10 steps, the transition probability showed less than 5\% deviation from the exact value.}
		\label{3_modes_fermion_Trotter_comparison_fig}
	\end{figure}

    To optimize the number of Trotter steps to be chosen, the exact evolution performed earlier was taken as a reference, and the Trotter evolution for values of $n_{\text{T}}$ ranging from $1$ to $10$ were performed and compared with it with the results shown in Fig.~\ref{Trotter_exact_comparison}. As the finite value of $n_{\text{T}}$ gives rise to an algorithmic error varying $\sim O(t^2/n_{\text{T}})$ \cite{trot_err_mitigate_1}, the trotterization results are seen to deviate significantly from the ideal results for larger values of the evolution time $t$. The circuit cost is decided by the choice of $n_{\text{T}}$ and we require a low circuit cost with reasonably low error. From Fig.~\ref{Trotter_exact_comparison} we see that up to  $t=0.2\, m_{\pi}^{-1}$ (green shaded region in the figure), the Trotter approximation yields results that match the exact evolution with a reasonable number of steps $n_{\text{T}}$ ranging from $5$ to $10$. Beyond this value of $t$ a greater number of Trotter steps that lead to prohibitive circuit costs are required and so we limit our numerical integration around $t=0.2\, m_{\pi}^{-1}$ or less. 
   
   From Fig.~\ref{exact_prob_vs_time}, we see that at $t=0.2 \, m_{\pi}^{-1}$, the probability of transition to the state $|f \rangle_1 \otimes |\phi\rangle_1$ is maximum at around 0.25. We are not only concerned with the approximation reproducing the transition probability accurately but also ensuring that forbidden transitions that do not conserve $K$ or $Q$ into other states are suppressed. The results of our trials using different Trotter steps are presented in Fig.~\ref{3_modes_fermion_Trotter_comparison_fig}. The rows correspond to different numbers of steps with the top row corresponding to the exact evolution. The columns label the various states that include the initial state, $|f\rangle_2$, as the first column and the final state, $|f \rangle_1 \otimes |\phi\rangle_1$, which is labeled as the one that conserves both $K$ and $Q$ as the second column. The last two columns collectively represent the remaining states grouped into those whose appearance in the simulation represents non-conservation of $K$ and $Q$ respectively (or both). The colors of each cell represent the probabilities of finding the respective state(s) at the end of $t=0.2 \, m_{\pi}^{-1}$. 

      We see from Fig.~\ref{3_modes_fermion_Trotter_comparison_fig} that if only one or two Trotter steps are used, significant probability leaks into states that do not conserve $K$ or $Q$. With 3 steps or more, there is no such noticeable probability leak, and with 7 steps or more the total probability of being either in the initial or final states (first two columns) was numerically equal to 1.0 in the simulations. The transition probability showed less than $10\%$ deviation from the exact value at 8 Trotter steps and this deviation dropped to less than $5\%$ by 10 steps. For the simple example considered, 10 Trotter steps correspond to a time step of $0.02 \, m_{\pi}^{-1}$. Based on this analysis and on considering other similar examples wherein the exact evolution is easy to compute, we chose the number of Trotter steps in subsequent simulations to be such that the time step is $0.02 \, m_{\pi}^{-1}$ or similar. Note that the simulations in both Figs. ~\ref{Trotter_exact_comparison} and ~\ref{3_modes_fermion_Trotter_comparison_fig} were done on the \textbf{ibmq\_qasm\_simulator}, provided by the IBM Quantum Experience \cite{IBMQ}, which supports up to 32 qubits and not on actual quantum hardware. The minimum value of $n_{\text{T}}$ that we arrived at is therefore set by the size and connectivity of the NISQ device and considerations of real noise, decoherence, etc. do not enter the picture at this stage.  
	
	\subsection{Varying the coupling constant} 
 
    As the next step, we look at the range of coupling constant $\lambda$ that can be explored given the limitations of the quantum devices we are working with. Studies of this kind require enormous amounts of resources to simulate efficiently, and hence, an exact evolution is not feasible. Here, we keep the total evolution time fixed at $t=0.2 \, m_\pi^{-1}$ with $n_{\text{T}} = 10$ and $N_{\text{max}}=4$. The number of Bosonic modals was also fixed at $3$. We considered four different types of initial states of the field here, namely, having one Boson alone in a particular mode of intermediate energy, having one Fermion alone in the similar energy mode, having one anti-Fermion alone in the similar mode, and a state in which one of each kind of particle is present in modes of similar energies. In the present case, the occupied mode for each case was chosen to be the $K = 2$ level, similar to the previous analysis on trotterization. In terms of qubit states these are  $|0000 \: 0000 \: 00 \, 01 \, 00\, 00\rangle$,  $|0100 \: 0000 \: 00 \, 00 \, 00\, 00\rangle$,  $|0000 \: 0100 \: 00 \, 00 \, 00\, 00\rangle$, and  $|0100 \: 0100 \: 00 \, 01 \, 00\, 00\rangle$, representing a single Boson, a single Fermion, a single anti-Fermion, and one particle of each kind, respectively.

    \begin{figure}[!htb]
		\centering
		\includegraphics[width=0.99\linewidth]{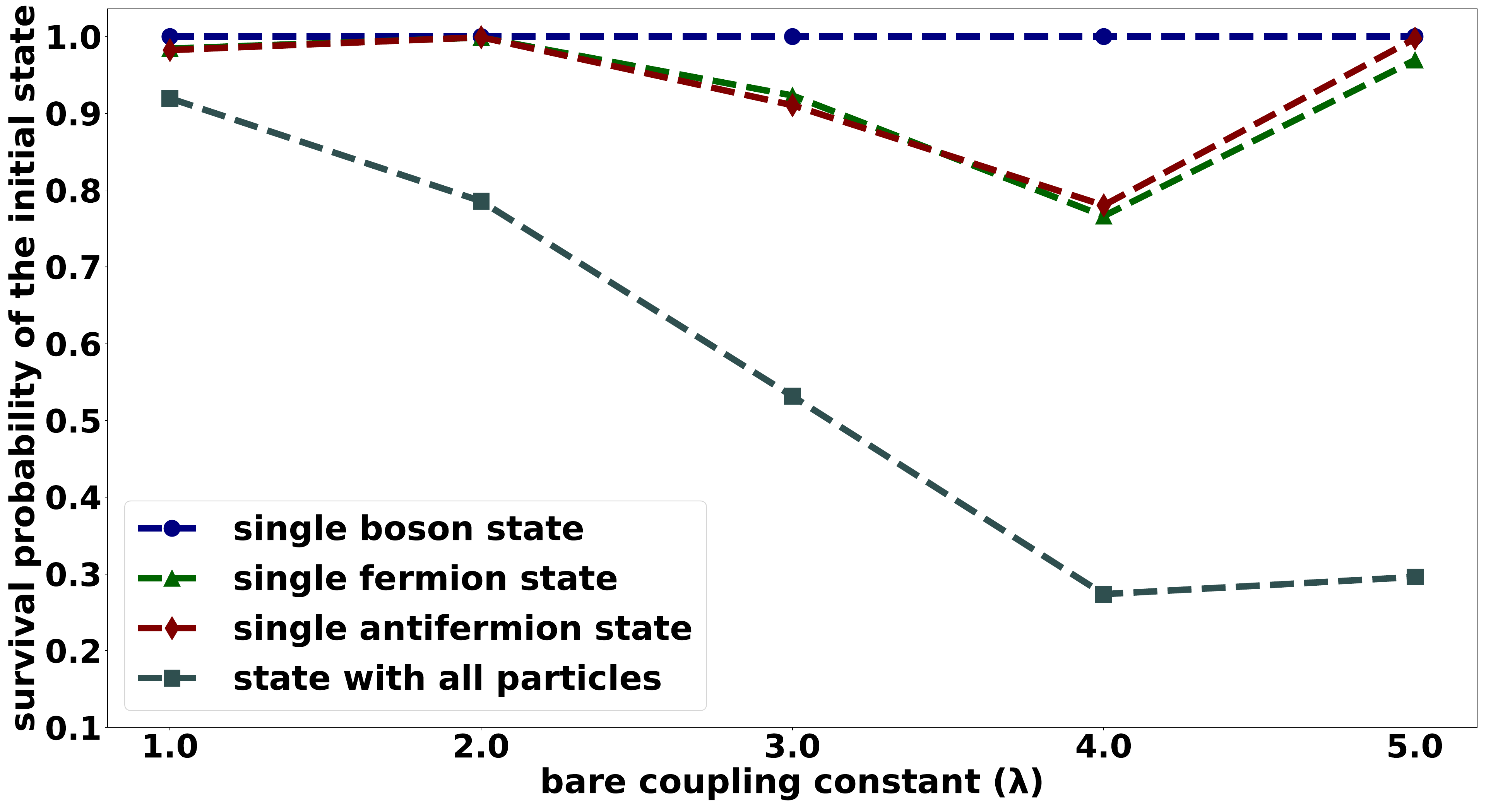} 
		\caption{Survival probabilities of the four initial states $|\phi\rangle_2 = |0000 \: 0000 \: 00 \, 01 \, 00\, 00\rangle$, $|f\rangle_2 =|0100 \: 0000 \: 00 \, 00 \, 00\, 00\rangle$, $|\bar{f}\rangle_2 = |0000 \: 0100 \: 00 \, 00 \, 00\, 00\rangle$ and $|f\rangle_2 \otimes |\bar{f}\rangle_2 \otimes |\phi\rangle_2 = |0100 \: 0100 \: 00 \, 01 \, 00\, 00\rangle$ as a function of the coupling constant $\lambda$. The total evolution time is $t=0.2\, m_\pi^{-1}$ with $n_{\text{T}} = 10$ and $N_{\text{max}}=4$ and number of Bosonic modals $3$. These results were obtained on the \textbf{ibmq\_qasm\_simulator}. }
		\label{Fig_L_comparison} 
	\end{figure}
 
	The simulations were done 8192 times
	each on the \textbf{ibmq\_qasm\_simulator}. The survival probability of each of the four initial states is plotted for different values of coupling in Fig.~\ref{Fig_L_comparison}.  The first initial state, $|\phi\rangle_2$ is an ``angel state" in the sense that it has no possible means of decaying into another state without violating the applicable conservation laws. Along expected lines, the survival probability of the angel state was observed to remain the same, equal to unity, for all values of coupling, as can be seen from Fig. \ref{Fig_L_comparison}. The remaining three states could evolve to other states, particularly for higher values of the coupling in the given time, The plots for the survival probability of the Fermion and anti-Fermion states (the second and third states considered) were observed to be nearly identical, owing to the fact that the conservation of the values of $K$ and $Q$ only permits a single kind of evolved state for both these cases, which involves the transition of the particle to a similar lower level with the emission of a Boson. The fourth state which includes all the three kinds of particles, has, as a result, a higher value of the harmonic resolution ($K = 6$), and hence quite a lot of valid output states are possible and observed. Significantly, even with several decay or transition channels available for this state, the probability for the appearance of $K$ non-conserving states was still minimal even for higher values of $\lambda$ considered. Thus, the value of $n_{\text{T}}$ chosen for these comparisons, based on the earlier analysis of the trotterization approximation, was figured to be good enough here (the case of $N_{\text{max}}=4$) as well, as can be seen from the results. The slight fluctuations in the survival probability of the initial state with no definite decreasing pattern at stronger coupling, can be inferred to be due to the finite, small-scale value of the number of modes. This established the range of $\lambda$ considered from 1 to 5 as being a parameter regime in which reliable results can be obtained through simulation using available NISQ devices. 

    \subsection{Varying $N_{\text{max}}$}

    While probability leaks to known forbidden states can be checked, one has to also check if allowed transitions to certain states are missed out because of the common limit $N_{\text{max}}$ placed on the number of modes for each field included in the simulation. This is more likely in the strong coupling regime than at weak coupling. We considered three values of $N_{\text{max}}$, $4$, $5$, and $6$ and investigated two values of the coupling constant, $\lambda = 1, 4$ corresponding to weak and strong couplings respectively. We set the total evolution time to $t=0.2 \, m_{\pi}^{-1}$ and we also considered the same four initial states as in the previous section. We find that irrespective of $\lambda$, probability amplitudes for new, allowed states do not appear on increasing $N_{\text{max}}$ for these initial states. The choice of $N_{\text{max}}=4$ made in the previous section is therefore a reasonable choice for computing evolution of these four initial states given the number of Trotter steps used and the required level of errors. We also find that when $N_{\text{max}}$ is increased the number of Trotter steps also has to be increased in order to keep the probability of leaking into forbidden states within acceptable limits. This is due to the appearance of more such states in addition to allowed states with an increase in $N_{\text{max}}$. Consequently, the circuit depth and cost increase substantially, indicating that larger values of $N_{\text{max}}$ are inaccessible to present-day NISQ devices. 
    
    When $N_{\text{max}}$ is increased, more levels become available above the occupied one. These levels can appear as intermediate states in various processes. However, including these levels also means an increase in the effective number of interaction terms in the Hamiltonian to be considered, leading to an increased circuit depth. Since the possible interactions have to confine the final states to the same sector of $K$ and $Q$, the single Boson and Fermion examples studied here that belong to the $N_{\text{max}} = 2$ sector do not have significant contributions from the higher levels. Since the final physical states we consider have to conserve $K$ and $Q$, no additional such state appears for the values of $K$ and $Q$ considered. However, for initial states belonging to sectors with larger values of $K$ or $Q$, the circuits will have to become larger, so as to incorporate more intermediate states and interactions, and hence would also require a larger number of Trotter steps to produce an approximation of the evolution with less errors. 
    This again would require more number of logical qubits, demanding more efficient error mitigation techniques to also be in place. Similar considerations apply to the case of increasing the number of Bosonic modals also. This was verified by increasing the number of the Bosonic modals from $3$ to $7$ by adding one more qubit to each Bosonic mode.

    \subsection{Simulating realistic processes}
    
	The systematic studies in the previous sections which were done on the \textbf{ibmq\_qasm\_simulator} provide confidence that the algorithm is working satisfactorily in the identified parameter regimes. Real validation is only possible through comparison with experiment. As noted earlier, without inserting realistic parameter values, we attempt to compare qualitative trends next. Quantitative agreement still remains beyond the scope of the devices that are accessible as seen, for instance, from the limitations on the order of the Suzuki-Trotter approximation, $N_{\text{max}}$, etc. Simulations can be done, at present, on an example state involving only a few light-front modes for each field, and that too in 1+1 dimensions only. The quantum simulation still allows computation of the time evolution of a physically reasonable initial state. Tracking this state in time allows for the computation of cross sections of specific processes that would be of interest in the relatively near future~\cite{roadmap_IBMQ} when the computation can be scaled up to 3+1 dimensions with more modes for each field.  

    The process of interest in the theory that we consider is the production of a pair of pions in a proton-proton collision: $p p \longrightarrow p p \pi^+ \pi^-$. Processes belonging to this type have been extensively studied in~\cite{cs_ref_1,cs_ref_2}. For our study, we chose $N_{\text{max}}=5$ with 3 modals for each Bosonic mode. Since our theory does not distinguish between the two kinds of charged Pions, they are effectively excitations of the same Bosonic field in the present case. A high value of the charged-pion coupling constant, $\lambda = 13.315$ similar to the values reported in \cite{coupling_const} was used here. The initial state was $|f\rangle_{4, 5} = |00011 \: 00000 \: 00 \, 00\, 00 \, 00\, 00\rangle$, with the protons occupying the $4^{th}$ and $5^{th}$ levels, belonging to the $K = 9$ sector of the theory. Time evolution of this state till $t = 0.4\, m_\pi^{-1}$ was computed keeping the time step low at a conservative value of $0.005 \, m_\pi^{-1}$ owing to the larger value of $K$. The simulation was done using the Qiskit Runtime Sampler primitive \cite{qkit_sampler}. 
	
	\begin{figure}[!htb]
		\centering
		\includegraphics[width=0.99\linewidth]{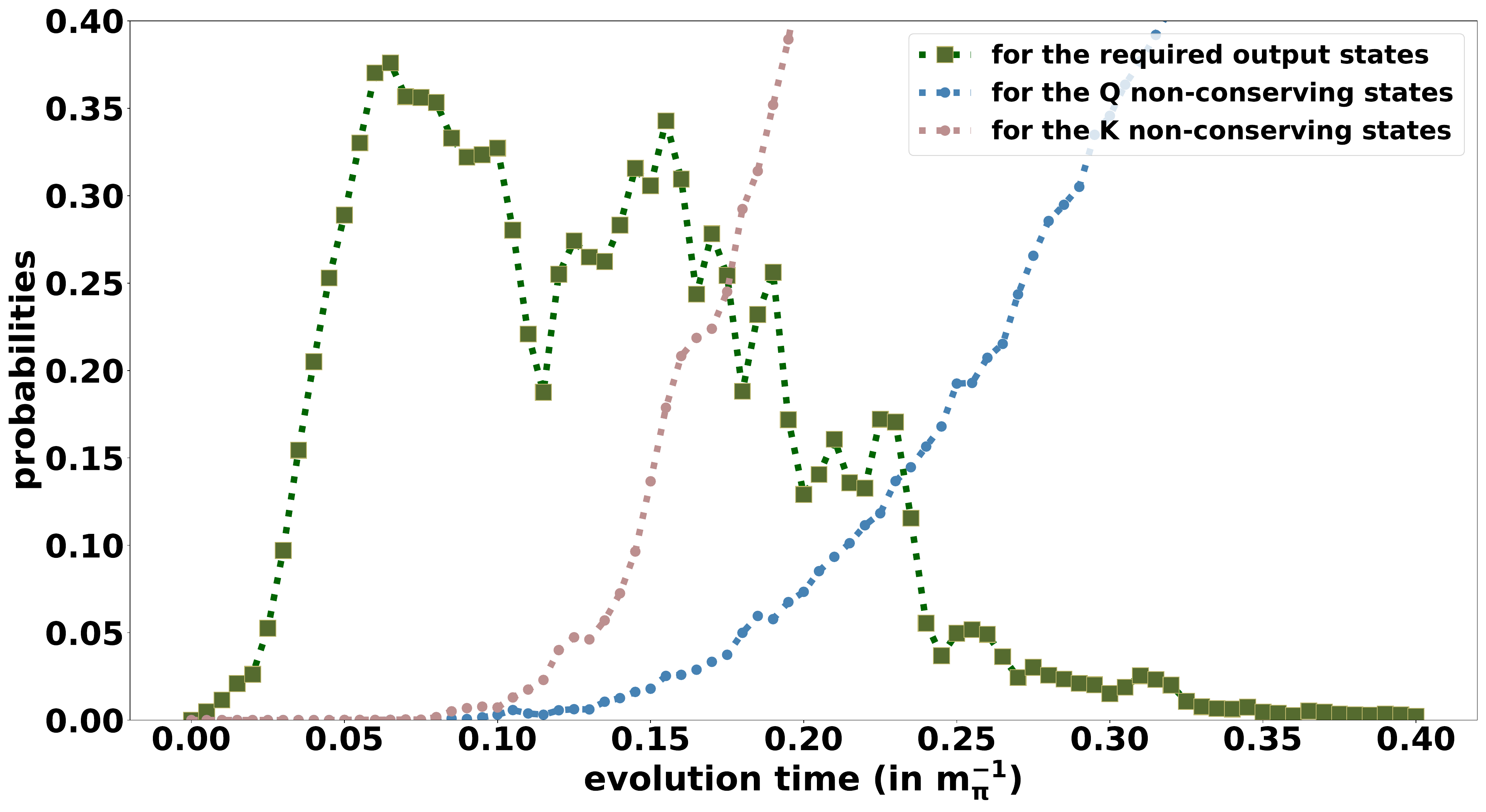} 
		\caption{The variation of the probabilities for the $p p \longrightarrow p p \pi^+ \pi^-$ process at different evolution times ranging from $0$ to $0.4$ $m_{\pi}^{-1}$, at timesteps of $0.005$ $m_{\pi}^{-1}$, for the initial state $|00011 \: 00000 \: 00 \, 00\, 00 \, 00\, 00\rangle$. The variation of the errors involving non-conservation of $Q$ and $K$, resulting from the trotterization approximation has also been shown here, with these errors increasing at larger values of the evolution times, as expected.}
		\label{prob_pp_collision}
	\end{figure}
	
     The total probability for obtaining any state in the same $K$-sector containing exactly two Bosons (Pions) and two protons (in order to conserve $Q$) was computed as a function of time, from the time-evolved state. The results are as shown in Fig. \ref{prob_pp_collision}. After the initial smooth rise in the values of the transition probability, which is expected, the Pion production probability appears to decay. The irregular nature of the decay suggests that the computation is not reliable beyond this point, as the probability is leaking to disallowed states and the computation itself is showing instabilities. This probability leak, due to the surge in the errors, resulting in $Q$ and $K$ non-conserving states, can be observed from the blue and pink lines respectively in Fig. \ref{prob_pp_collision}. The finite small-scale value of $n_{\text{T}}$ used in these simulations for all values of the evolution time $t$ is the major contributor to this error, which scales as $O(t^2/n_{\text{T}})$. Even though these results are not dependable for these longer values of evolution time, the trend of the steady increase in the probabilities of the required state and the beginning of its decline after reaching a peak (till around $t = 0.1 \, m_{\pi}^{-1}$) is a feature that is along expected lines. It may be noted that after a short transient, the transition rate (time derivative of the transition probability) remains constant over an interval of time allowing comparison with available measured rates. Fixing the interaction time based on observed times will allow computation of the scattering cross sections as well. However, typical results in the literature as in~\cite{cs_ref_1} \cite{cs_ref_2} reports measured cross sections and rates corresponding to different combinations of Pion types corresponding to processes like $p p \longrightarrow p p \pi^+ \pi^-$, $p p \longrightarrow p p \pi^0 \pi^0$, $p p \longrightarrow p n \pi^+ \pi^0$, etc. Comparison with these results requires expansion of the theory to include multiple, distinct Bosonic fields which is beyond the scope of the current work. Also, similar analyses for different initial states of this kind need to be performed on a larger scale, so as to obtain valid and appropriate values for these probabilities for initial states with different energies. Furthermore, the quantum simulation provides the additional benefit of accessing superposition states of different proton pair occupancies as initial states so as to access the intermediate energy values that do not correspond to any specific energy level. These values can then be transformed back to the equal-time frame to compare with the available experimental values for verification, and later on with advances in the resources available, can be utilized for making reasonable predictions for particle collision experiment results.  
     
    \subsection{Simulations on real quantum hardware}
	
	Finally, the performance of the simulation on an actual quantum hardware remains to be examined. A minimal example was chosen and run on the \textbf{ibm\_perth} processor, a 7-qubit processor of the Falcon type of processors from IBM Quantum. This processor has a QV (Quantum Volume) value of 32 and CLOPS (circuit layer operations per second) value of 2.9K, and hence can almost perfectly run circuits of width and depth each 5, respectively. An example with only two modes for each field (i.e. $N_{\text{max}}$ = 2) was considered and each mode of the Bosonic field was set to have only one modal, making it effectively identical to a Fermionic mode. The initial state considered was of a single proton alone in the second level, represented in the Fock space as $|01 \: 00 \: 00\rangle$, which is similar to the sample states considered in the previous sections. The simulations were done for two values of $\lambda$ $(= 1.0 \text{ and } 4.0)$ and for a fixed evolution time of $t = 0.2 \, m_{\pi}^{-1}$. The Hamiltonian was approximated by excluding the terms quadratic in $g$, i.e., $H_S$ and $H_F$ (See Appendix \ref{appA}), thereby limiting the interactions to be of the vertex type ($H_V$) alone, giving $H = H_M + H_V$. This is justifiable, since the only allowed interaction for the state considered is governed by the vertex part of the Hamiltonian. 

    \begin{figure}[!htb] 
        \centering
        \subfloat[The results for the case of $\lambda = 1.0$.]{%
\includegraphics[width=0.99\linewidth]{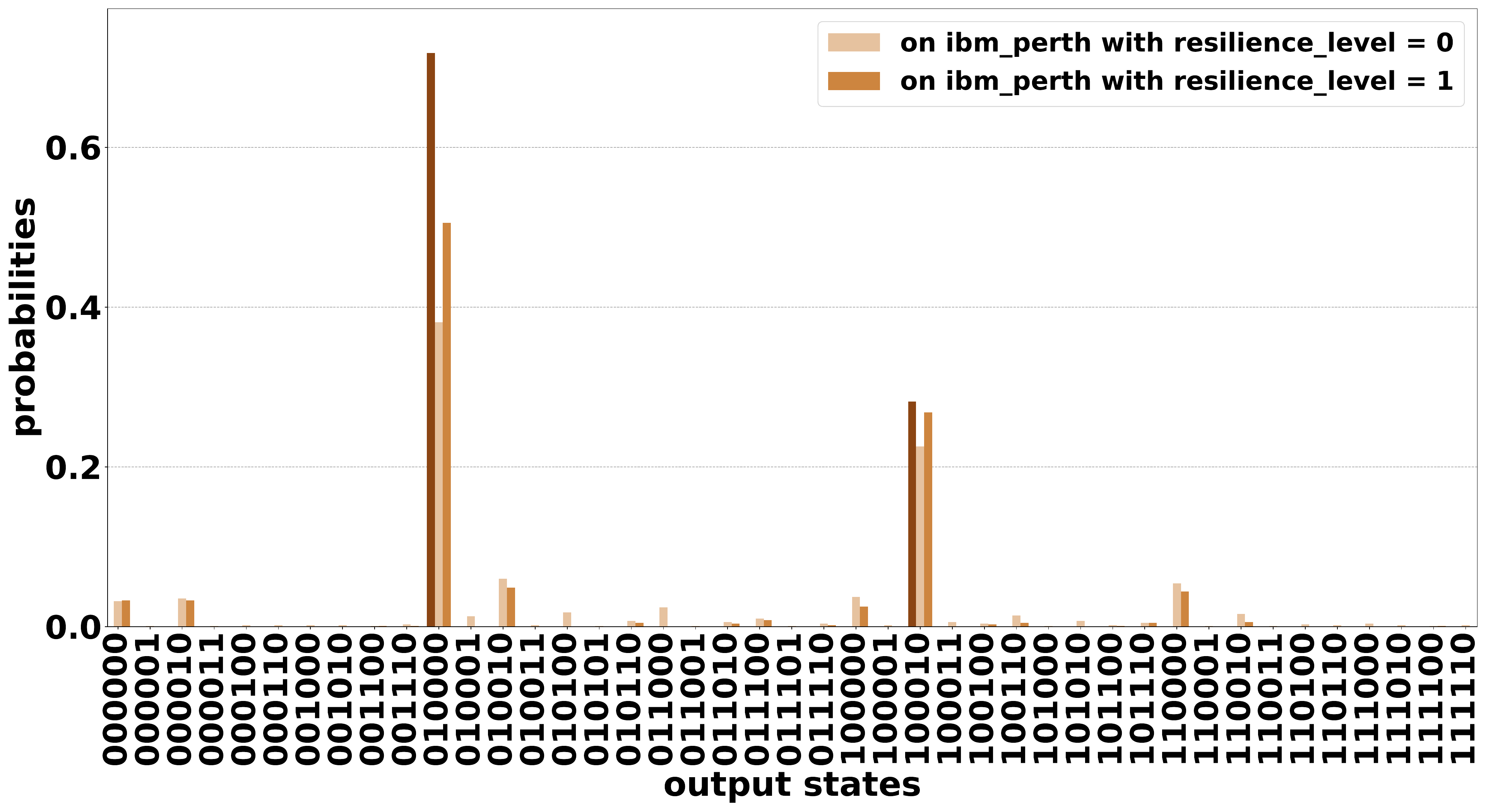}\label{evol_with_time_1.0}}
        \hfill
        \subfloat[The results for the case of $\lambda = 4.0$.]{%
\includegraphics[width=0.99\linewidth]{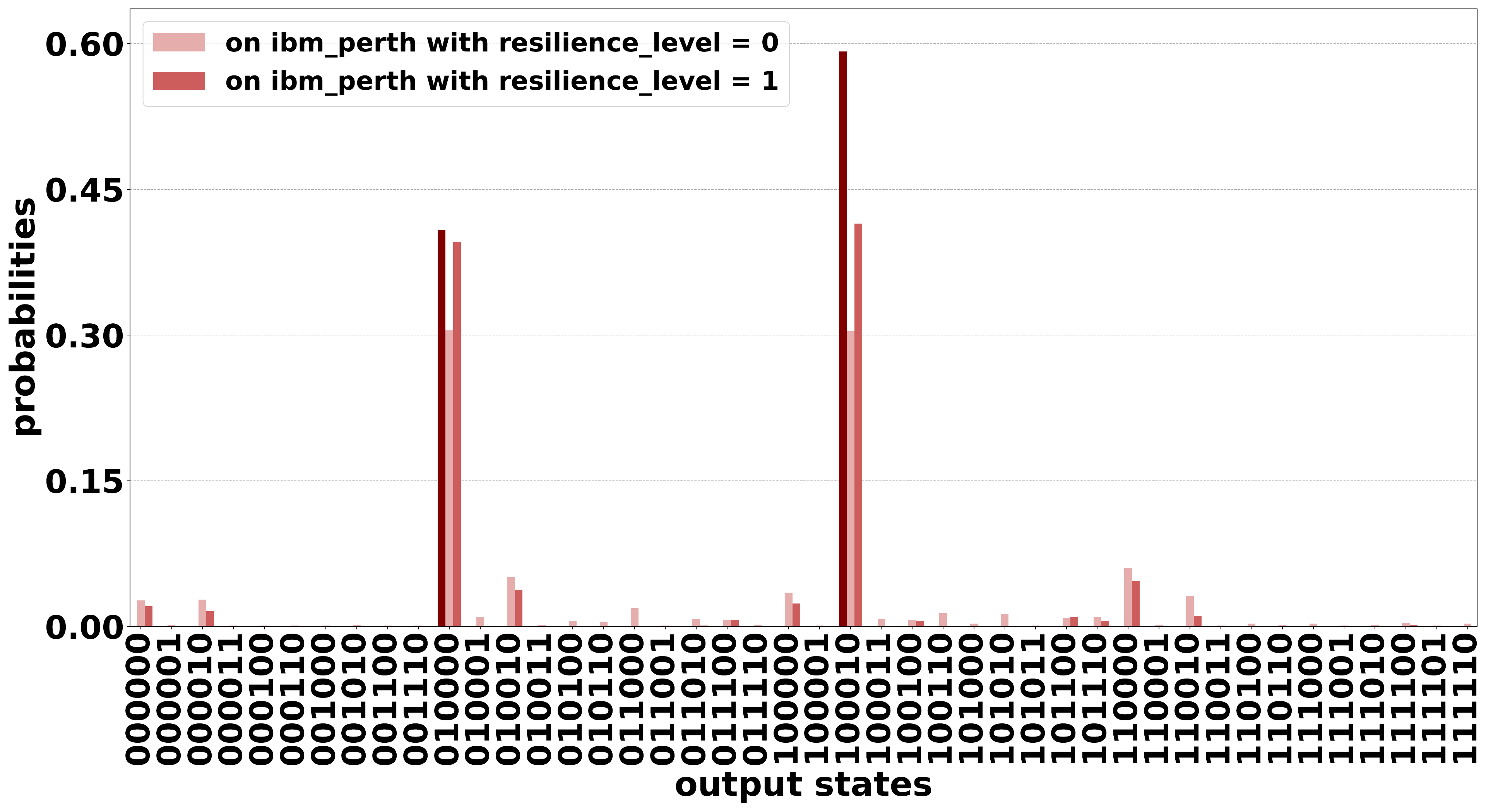}\label{evol_with_time_4.0}}
    \hfill
    \caption{The results of a simple instance of a proton-pion field evolution on both the \textbf{ibmq\_qasm\_simulator} and the real hardware \textbf{ibm\_perth} for two different values of the coupling constant $\lambda$ (The program runs were done on September $11$, $2023$). The histograms shown with the darker shade in the first and second distributions denote the results from the simulator run while those shown in the lighter shades correspond to results from the \textbf{ibm\_perth} processor. We see that there is good agreement between the two. The simulation parameters used however are very far from being realistic due to the constraints placed by the small size of the available quantum hardware.}
    \label{dev_run_results} 
    \end{figure}

    \begin{figure}[!htb] 
        \centering
        \subfloat[The errors for the case of $\lambda = 1.0$.]{%
\includegraphics[width=0.99\linewidth]{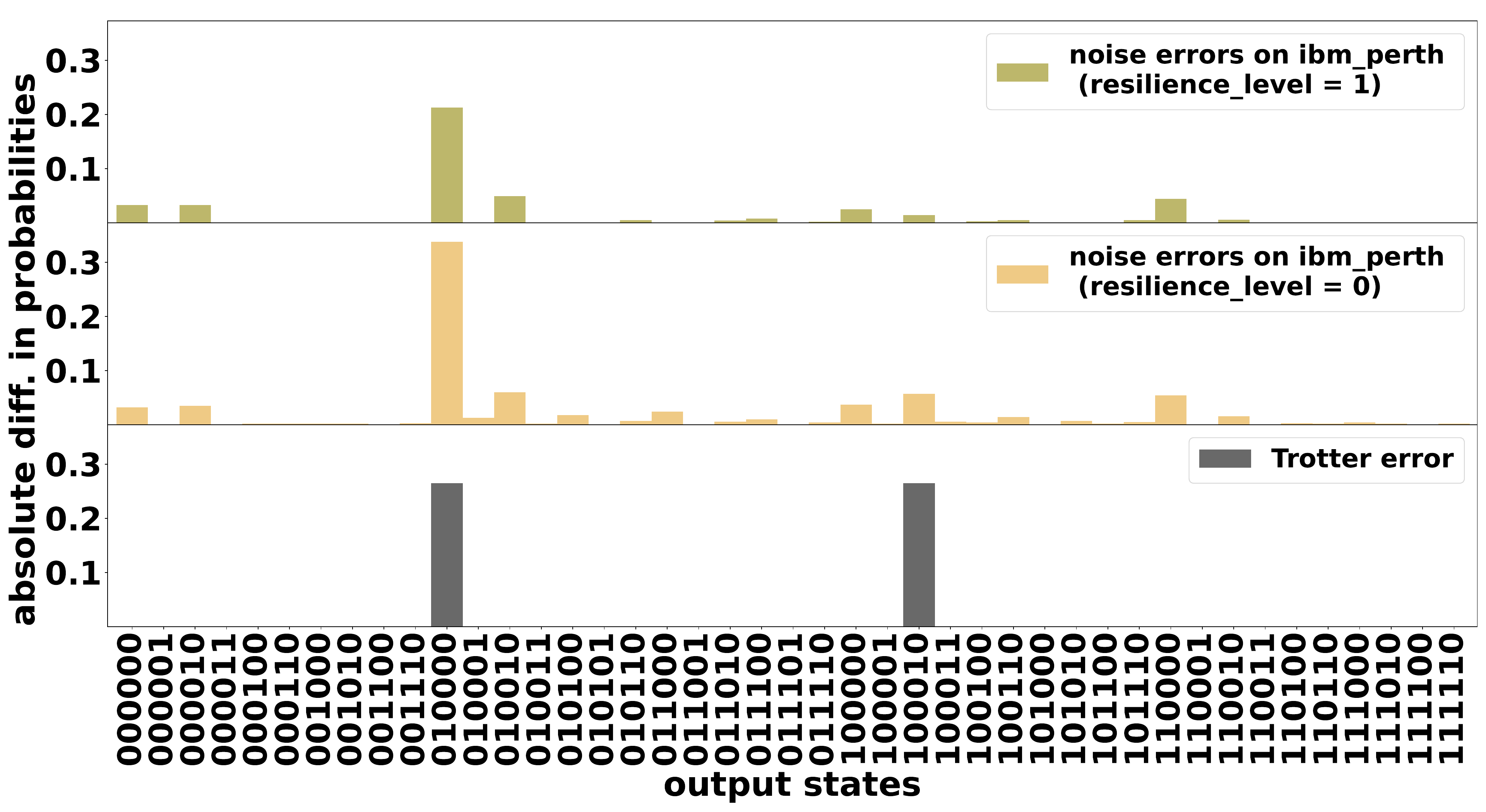}\label{err_on_dev_1.0}}
        \hfill
        \subfloat[The errors for the case of $\lambda = 4.0$.]{%
\includegraphics[width=0.99\linewidth]{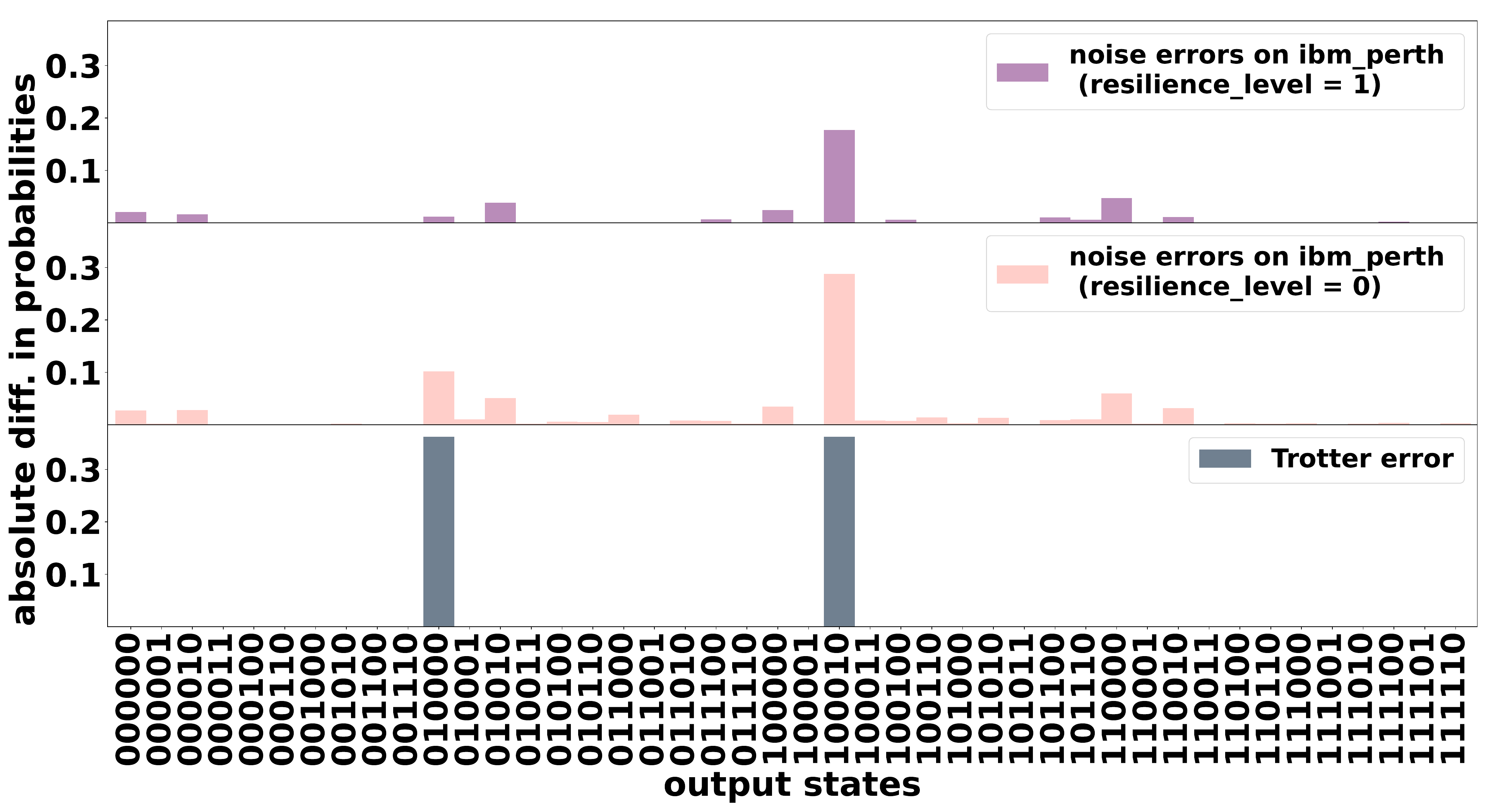}\label{err_on_dev_4.0}}
    \hfill
    \caption{Characterization of the errors involved in the device runs illustrated in Fig. \ref{dev_run_results}. The trotterization errors were obtained by comparing the simulator results with the exact evolution results, while the errors due to the hardware noise were obtained by comparing the device run results with the results from the simulator. Absolute values of the deviation of the probabilities are plotted here.}
    \label{dev_run_errors} 
    \end{figure}
    
    On the \textbf{ibmq\_qasm\_simulator}, with $8192$ shots of the execution, and a single Trotter step for the evolution, this state showed a transition of the proton to the $K = 1$ level, resulting in a Pion being produced in the lowest energy mode of the Bosonic field, i.e., the state $|10 \: 00 \: 10 \rangle$, with a probability of $\sim 0.28$, for $\lambda = 1.0$. This transition probability showed an increase to $\sim 0.6$ for $\lambda = 4.0$ on the simulator. The limitations of the actual quantum device are quite severe and so we were able to do only a single Trotter step. Accordingly, the calculations done for comparison on the simulator were also restricted to a single step, which, as we have already seen, is not ideal. 
    
    For the execution on the device with the same number of shots, the circuit was transpiled into one matching with the coupling map and the basis gates specific to the device, with a noise-adaptive layout method. For both the values of the coupling constant, this circuit had a depth of $\approx 45$ and with no noise mitigation techniques used, this did give results similar to the simulator run, although with some small probabilities for various other output states, which can be ascribed to the noise in the device. A slight improvement in the performance, in terms of the reduction of this noise, was obtained with the noise resilience\_level set to $1$ for the device run, which utilizes some pre-defined noise-mitigation procedures. The results of all these $3$ runs for both the values of $\lambda$ examined here can be seen in Fig. \ref{dev_run_results}. The errors arising from the Trotter approximation were determined by comparing the simulator run results with the results of the exact evolution for this example. The simulator run does not result in any incorrect output states. The two output states observed are the same as those from the exact evolution, with the Trotter approximation only affecting the relative probability of these two states. The absolute values of deviations of the probabilities obtained from the simulator from those corresponding to the exact evolution are represented as the trotterization error in Fig.~\ref{dev_run_errors}. The errors resulting from the noise in the hardware were computed by similarly comparing the device run results with the simulator run results and evaluating the absolute differences which are also shown in Fig.~\ref{dev_run_errors}. Since the simulator run already includes the Trotter errors, these differences would correspond to the errors arising from the noise in the device alone. Slight probability leaks into invalid states are observed in this case. We see that there is close agreement between the actual runs and the ones on the simulator, giving confidence that as the size of the hardware becomes larger, it would indeed be able to address field theoretic computations of relevance to modern-day particle physics on gate-based quantum computers. 
	
	\section*{Discussion and Conclusion}

    We have demonstrated in detail how a simple quantum field theory can be simulated on a gate-or-circuit-based quantum computer. The key to regulating the number of states to be considered in such a simulation on a finite-sized quantum computer was using the light-front formulation of the theory. Our focus has been on keeping the details involved in the theoretical formulation as minimal as possible so that the methodology for setting up such a simulation is clear. This comes at the price of forsaking the appropriate units, scales and finer detail that would have enabled a direct comparison of the results of the simulation with experimental results. We also notice that the level of detail that can practically be included in NISQ devices that are currently available is bound to be quite insufficient for such direct comparison in any case. We have laid out the procedure for truncating the size of the Hilbert space required and examined the inaccuracies introduced by the truncation as well as the trotterization procedure that is required to compute the unitary time evolution in the given field theory. We see that even for the simple field theory we consider in 1+1 dimensions, the cost in terms of circuit size and depth can become prohibitive quite rapidly as we explore initial states of the theory with higher and higher total energies. We have established the limits of the number of excitations that can be included in a simulation of a particular size and the largest size of each trotter step that can be considered before the simulation results start deviating substantially from the exact ones. 

    We also introduced a systematic construction for mapping the Bosonic operators that appear in the theory to strings of Pauli operators that act on the qubit register in the quantum computer. We tested out our theory on the \textbf{ibmq\_qasm\_simulator}. To serve as a representative of the initial states and methods involved in computing the cross sections of various possible processes in a particle collision experiment, a modest example with only $5$ modes for each kind of particle was studied, and the probabilities of a particular output state were recorded. This, when performed on a larger scale, can aid in effectively calculating reaction cross sections for collisions happening in particle accelerators. Finally, a small-scale example was chosen to demonstrate the performance of this study of the dynamics of the theory, on a real NISQ hardware - the 7-qubit \textbf{ibm\_perth} processor. Even though with the present advances in the technology, only a minimal example of this kind could be successfully carried out, it indeed serves as a promise for the viability of the algorithm with more resources and error mitigation techniques in place. Optimizing the various stages of the simulation with specific error mitigation techniques, both pre-defined as well as custom ones, in order to make the algorithm suitable for large-scale applications still remains to be done as future work. Alternate mapping methods, like compact mapping, or other novel techniques for the same, can also be put to use to achieve this optimization goal. These extensions hold the promise of making our present work suitable for practical use cases and computation of experimentally measurable cross sections, decay rates, etc. Extension of our approach to $(3+1)$ dimensional field theories is also another challenging frontier that remains to be explored. 

    \appendix*
    
    \section{$H$ in the light-front frame}
    \label{appA}

   The Hamiltonian or Energy Operator in the light front frame for the theory we consider has four terms,
	\begin{equation}
		\label{H}
		H = H_M+H_V+H_S+H_F,
	\end{equation} 
	The first term, $H_M$, is the mass term and it is given by,
    \begin{align*}
			H_M & = & \!\!\!\!\! \sum_n \frac{1}{n}[a_n^\dagger a_n (m_B^2 +  g^2 \alpha_n) + b_n^\dagger b_n (m_F^2 + g^2 \beta_n) \nonumber \\
			& &   + \, d_n^\dagger d_n(m_F^2 + g^2 \gamma_n)],
	\end{align*}
    which depends on the bare Fermion and Boson masses and also on the self-induced inertias ($\alpha$, $\beta$, and $\gamma$). The self-induced inertias are pure numbers that allow one to choose $k^+$ and $k^-$ to be positive for all states that are considered (see Fig.~\ref{fig1}) and are given by,
	\begin{align*}
			\alpha_n  = & \sum_{m=1}^{\Lambda} (\{n-m|m-n\} - \{n+m|-m-n\}), \\
			\beta_n  = & \sum_{m=1}^{\Lambda} \frac{n}{m} \{n-m|m-n\}, \\
			\gamma_n  = & \sum_{m=1}^{\Lambda} \frac{n}{m} \{n+m|-m-n\},
	\end{align*}
    where
    \begin{equation}
		\label{nm}
		\{n|m\} = \begin{cases}
			0 &\text{ if } n = 0 \text{ or } m = 0, \\
			\frac{1}{n}\delta_{m,-n} &\text{ otherwise } \\
		\end{cases}.
	\end{equation}     
    The other three terms in the Hamiltonian, namely the vertex part ($H_V$), the seagull part ($H_S$), and the fork part ($H_F$), with their names attributed to their graphical representations \cite{Pauli_Brodsky_1993}\cite{Pauli_Brodsky_2001}, are the ones involving the interactions.  These contain interaction terms that are cubic and quartic in the creation and annihilation operators respectively \cite{light_front}. The $H_V$ part is linear in the coupling constant $g$, while the $H_S$ and $H_V$ are quadratic in $g$, where $g \equiv \lambda/\sqrt{4\pi}$~\cite{Pauli_Brodsky_2001}. The interaction terms are given by, 
	\begin{align*}
		H_V   =  &  g m_F  \sum_{k,l,m} \Big[  \\
        & (b_k^\dagger b_m c_l^\dagger + b_m^\dagger b_k c_l) (\{k+l|\!-\!m\} + \{k|\!+l\!-\!m\}) +  \\ 
        &  (d_k^\dagger d_m c_l^\dagger + d_m^\dagger d_k c_l)  (\{k+l|\!-\!m\} + \{k|\!+\!l\!-\!m\}) - \\ 
        & (b_k d_m c_l^\dagger + d_m^\dagger b_k^\dagger c_l)  (\{k-l|\!+\!m\} + \{k|\!-\!l\!+\!m\}) \Big],
	\end{align*} 
	\begin{align*}
		H_S  = &  g^2 \sum_{k,l,m,n} \Big[ &  \\
        &  b_k^\dagger b_m c_l^\dagger c_n (\{k-n|l-m\} + \{k+l|-m-n\}) \\
        & + d_k^\dagger d_m c_l^\dagger c_n (\{k-n|l-m\} + \{k+l|-m-n\}) \\ 
        & + (d_k b_m c_l^\dagger c_n^\dagger + b_m^\dagger d_k^\dagger c_n c_l) \{l-k|n-m\} \Big],
	\end{align*} 
	\begin{align*}
			H_F  = &  g^2  \sum_{k,l,m,n} \Big[ & \\
        & (b_k^\dagger b_m c_l^\dagger c_n^\dagger + b_m^\dagger b_k c_n c_l) \{k+l|n-m\} \\ 
        &+ (d_k^\dagger d_m c_l^\dagger c_n^\dagger + d_m^\dagger d_k c_n c_l) \{k+l|n-m\} \\ 
        &+ b_k^\dagger d_m^\dagger c_l^\dagger c_n (\{k-n|m+l\} + \{k+l|m-n\}) \\ 
        &+ d_m b_k c_n^\dagger c_l (\{k-n|m+l\} + \{k+l|m-n\})\Big],
	\end{align*}
	where $c_n = a_n/\sqrt{n}$. More details on light-cone-quantized QCD in $(1+1)$D can be found in \cite{DLCQ_1, DLCQ_2} and a detailed review of the quantization procedure for different field theories on the light cone in \cite{review_DLCQ}.
	
\begin{acknowledgements}
This work was supported by the QuEST program of the Department of Science and Technology through project No. Q113 under Theme 4. We acknowledge the use of IBM Quantum services for this work. The views expressed are those of the authors, and do not reflect the official policy or position of IBM or the IBM Quantum team. Gayathree M. Vinod would also wish to thank Dr. Tanumoy Mandal, Assistant Professor at IISER Thiruvananthapuram, former IISER student Kartik Bhide, and one of the doctoral students at IISER - Rachit Sharma, for helping in understanding certain terminology specific to experimental particle physics. She also likes to thank all the participants of the Frontier Symposium in Physics (2023) organized by IISER Thiruvananthapuram, for their relevant queries related to this work.
\end{acknowledgements}
	
\bibliography{ref_qft_simulation}

\end{document}